\documentclass[preprint]{aastex62}
\usepackage[T1]{fontenc}
\usepackage{microtype}
\usepackage{listings}
\usepackage{afterpage,amsmath,amssymb,graphicx,natbib, color}
\usepackage{hyperref}
\usepackage{array}   

\newcommand\acronym{\texttt{ACRONYM}}
\begin{document}
\title{ACRONYM: Acronym CReatiON for You and Me}
\author{B.~A.~Cook}
\affiliation{Center for Astrophysics | Harvard \&{} Smithsonian, 60 Garden St., Cambridge, MA 02138, USA}
\correspondingauthor{B.~A.~Cook}
\email{bcook@cfa.harvard.edu}

\begin{abstract}
    Each year, countless hours of productive research time is spent brainstorming creative acronyms for surveys, simulations, codes, and conferences.
    We present \acronym{}, a command-line program developed specifically to assist astronomers in identifying the best acronyms for ongoing projects.
    The code returns all approximately-English-language words that appear within an input string of text, regardless of whether the letters occur at the beginning of the component words (in true astronomer fashion).
\end{abstract}

\section{Introduction}
\label{s.intro}
The field of astronomy is replete with \textit{acronyms}: abbreviations formed from individual letters within a longer word or phrase (the \textit{compound term}).
Everything from journals (\texttt{ApJ}, \texttt{MNRAS}) to observatories (\texttt{HST}, \texttt{LIGO}, \texttt{SDSS}) to simulations (\texttt{FIRE}, \texttt{EAGLE}) are consistently given long, detailed names that are later shortened for ease of daily use\footnote{While abbreviations such as \texttt{HST} that are not pronounced as a single word are technically \textit{initialisms}, Merriam-Webster concludes the term acronym can be used interchangeably to refer to both categories.}.

Grammatically speaking, acronyms are usually constructed from the first letter of each word in the compound term, with articles, prepositions, and coordinating conjunctions (such as \textit{a}, \textit{and}, and \textit{in}) occasionally excluded.
Compared to other scientific fields, astronomers seem particularly comfortable with generating forced acronyms, formed from any ordered subset of letters within the compound term, preferring ease of pronunciation over strict adherence to the traditional rules of acronym decorum. Extensive web-pages and blog posts have been dedicated to cataloguing particularly unusual astronomical acronyms, perhaps none as famous as Glen Petitpas' \texttt{DOOFAAS} (Dumb Or Overly Forced Astronomical Acronyms Site)\footnote{\url{https://www.cfa.harvard.edu/~gpetitpas/Links/Astroacro.html}}. Of the many listed examples, a few notable ones include \texttt{H0LICOW} (H$_0$ Lenses In COSMOGRAIL's Wellspring), \texttt{BATMAN} (BAsic Transit Model cAlculatioN in python), and \texttt{ENIGMA} (EvolutioN of Grains in the MAgellanic clouds).
As a group, astronomers are likely surpassed only by U.S.~lawmakers in their love of convoluted acronyms.

In this note, we present Acronym CReatiON for You and Me, or \acronym{}, a command-line code that identifies English acronyms from any arbitrary string of text.
All "astronomy-valid" acronyms, as well as the appropriate capitalization of the input phrase, are returned to the user, sorted with longest (and therefore best) acronyms first.
With widespread use, \acronym{} should revolutionize the field of astronomy by freeing researchers from the burdensome task of brainstorming convoluted acronyms for their projects, adding back countless hours of productive work to the research cycle.
\pagebreak
\section{Code Installation and Use}
\label{s.code}

\acronym{} is an open-source Python package, hosted on GitHub (\url{https://github.com/bacook17/acronym}) and available through the Python Package Index (\texttt{PyPi}). It can be installed from the command line in a single line using Python pip:
\begin{lstlisting}
$ pip install acronym
\end{lstlisting}

Installing \acronym{} makes available a command-line program by the same name that allows acronyms to be generated for any input string of alpha-numeric text.
The program is executed by passing a string of text, enclosed within quotes.
The resulting acronym suggestions are printed to standard-output, or can optionally be saved to a specified file.

An example, along with the first few resulting suggestions, is as follows:
\begin{lstlisting}
$ acronym "the long name of your very fancy project"
Collecting word corpus
Identifying matching acronyms
Process Complete
TERRACE	ThE long name of youR veRy fAnCy projEct
THEREAT	THE long name of youR vEry fAncy projecT
TYRRANY	The long name of YouR veRy fANcY project
TAVERN	The long nAme of your VERy faNcy project
TEAPOT	ThE long nAme of your very fancy PrOjecT
TENANT	ThE loNg name of your very fANcy projecT
TENOUR	ThE loNg name of yOUr very fancy pRoject
TENURE	ThE loNg name of yoUr very fancy pRojEct
[...]
\end{lstlisting}

The \acronym{} algorithm begins by collecting all unique English words from a corpus provided by the Natural Language Toolkit \footnote{\url{https://www.nltk.org/}}.
Several corpora are available to choose from: the default is the general \textit{Words} corpus. Users may prefer to use a more strict corpus, containing more commonly-used English words, and these can be chosen through the "-s" (\textit{Brown}) or "-ss" (\textit{Gutenberg}) flags.

The algorithm then selects all words from the corpus that begin with the same letter as the input phrase (the only restriction we place on acronyms is they begin with the same letter).
It then examines each word to see whether it is a valid acronym by recursively searching for the first and last letters of the potential acronym from within the input phrase. By default, \acronym{} searches only for acronyms of between 4 and 8 letters, but additional command-line arguments allow for customizing these values.

A single acronym could potentially have multiple realizations within the same phrase (such as \texttt{THEM} for "The study of Helium ExperiMentally" and "THe study of hElium experiMentally").
The algorithm will only identify a single such instance, and no guarantee is provided that it will be the ideal matching acronym.

\section{Future Work}
\label{s.future}

The \acronym{} algorithm is brute-force, manually checking all words from a given corpus (that begin with the same first letter) against the input phrase. 
This makes the specified corpus of words the limiting factor in identifying good acronyms, and no corpus is perfect for this job.
Allowing for integration of corpora from languages other than English would be extremely straightforward, and would be a significant first step towards increasing the accessibility of the code.

The execution time of the code also scales nearly linearly with the size of the chosen corpus, restricting the complexity of word combinations that can be attempted.
Adding combinations of short words (such as \texttt{KINGFISH}) to the corpora would open many new possibilities for good acronyms, but the number of such combinations makes the approach intractable.

As of this publication, \acronym{} returns a single list of all identified acronyms, sorted by length and then alphabetically, requiring the user to then manually identify the preferred acronyms.
Returning results sorted by some sort of scoring metric would make the experience easier for users.
What exact metric defines a "good" acronym is difficult to judge, but factors could potentially include the number of first letters from the component words used, the overall length of the acronym, and the frequency of the acronym's common use.

Finally, the resulting acronym is restricted to use only the phrase as exactly entered, potentially missing a perfect acronym that could result from reversing the order of two words in the phrase or replacing a word with a synonym. 
Altering the name of a project in such a way as to find the ideal acronym is far outside the scope of the code, and is delegated to the user.
Future advances in AI should hopefully bring automated project naming\footnote{not to mention automated data analysis, data collection, grant writing, talk giving, and general science-doing} into the realm of possibility.

\acknowledgments
The author thanks Harshil Kamdar, James Guillochon, Lehman Garrison, and Sebastian Gomez for thoughtful suggestions and running \acronym{} through useful tests.

\end{document}